# A wide-range survey on Recall-Based Graphical User Authentications algorithms based on ISO and Attack Patterns

**ARASH HABIBI LASHKARI**
Computer Science and Information Technology,
University of Malaya (UM)
Kuala Lumpur, Malaysia .

**SAMANEH FARMAND**
Computer Science and Information Technology (IT),
University Malaya (UM)
Kuala Lumpur, Malaysia .

**DR. ROSLI SALEH**
Computer Science and Information Technology,
University of Malaya (UM)
Kuala Lumpur, Malaysia .

**Dr. OMAR BIN ZAKARIA**
Computer Science and Data Communication (MCS),
University of Malaya (UM)
Kuala Lumpur, Malaysia
.

**Abstract- Nowadays, user authentication is one of the important topics in information security. Text-based strong password schemes could provide with certain degree of security. However, the fact that strong passwords being difficult to memorize often leads their owners to write them down on papers or even save them in a computer file. Graphical user authentication (GUA) has been proposed as a possible alternative solution to text-based authentication, motivated particularly by the fact that humans can remember images better than text. In recent years, many networks, computer systems and Internet-based environments try used GUA technique for their user's authentication. All of GUA algorithms have two different aspects which are usability and security. Unfortunately, none of graphical algorithms were being able to cover both of these aspects at the same time. This paper presents a wide-range survey on the pure and cued recall-based algorithms in GUA, based on ISO standards for usability and attack patterns standards for security. After explain usability ISO standards and attack patterns international standards, we try to collect the major attributes of usability and security in GUA. Finally, try to make comparison tables among all recall-based algorithms based on usability attributes and attack patterns those we found.**

.

## I. INTRODUCTION

In recent years, computer and network security has been formulated as a technical problem. A key area in security research is authentication which is the determination of whether a user should be allowed access to a given system or resource. In this respect, the password is a common and widely authentication method still used up to now.

The use of passwords goes back to ancient times when soldiers guarding a location by exchange a password and then only allow a person who knew the password. In modern times, passwords are used to control access to protect computer operating systems, mobile phones, auto teller machine (ATM) machines, and others. A typical computer user may require passwords for many purposes such log in to computer accounts, retrieving e-mail from servers, accessing to files, databases, networks, web sites, and even reading the morning newspaper online. In graphical password, the problem arises because passwords are expected to have two fundamentals requirements:

i. Password should be easy to remember.

ii. Password should be secured.

In a graphical password system, a user needs to choose memorable image. The process of choosing memorable images depends on the nature of the process of image and the specific sequence of click locations. In order to support memorize ability, images should have meaningful content because meaning for arbitrary things is poor.

## II. RECALL-BASED ALGORITHMS

A literature on most of articles regarding graphical password techniques from 1994 till 2009 shows that graphical password can be categorized into pure recall, cute recall and recognition groups. This section tries to explain Pure and cute Recall-Based algorithms by focusing on their lacks and weaknesses.

### 1. Passdoodle (Pure recall)

Passdoodle is a graphical password comprised of handwritten designs or text, usually drawn with a stylus onto a touch sensitive screen. In their 1999 paper, Jermyn et al. prove that doodles are harder to crack due to a theoretically much larger number of possible doodle passwords than text passwords [18] . Figure 1 will be shown a sample of Passdoodle password.

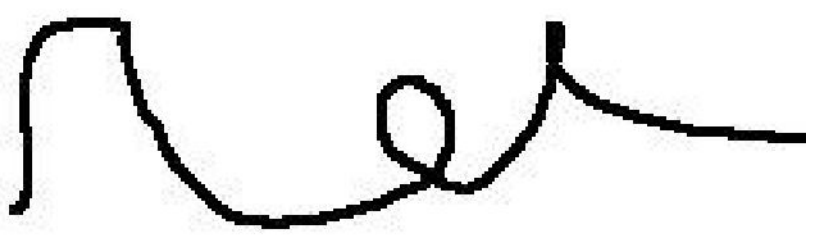

Figure 1: An Example of a Passdoodle

***Weaknesses***: with reference to the [18], they found that people could remember complete doodle images as accurately as alphanumeric passwords, but they were less likely to recall the order in which they drew a doodle than the resulting image.

In the other research [24], users were fascinated by the doodles drawn by other users, and frequently entered other users' login details merely to see a different set of doodles from their own.

### 2. Draw A Secret (DAS) (Pure recall)

In 1999, this method present by allowing the user to drawing a simple picture on a 2D grid as in Figure 2. The interface is consisting of a rectangular grid of size G * G. Each cell in this grid is denoted by discrete rectangular coordinates (x,y). As it can be seen in the figure, the coordinate sequence generated by drawing is [7]: (2,2), (3,2), (3,3), (2,3), (2,2), (2,1), (5, 5). Figure 2 shows a sample of DAS password.

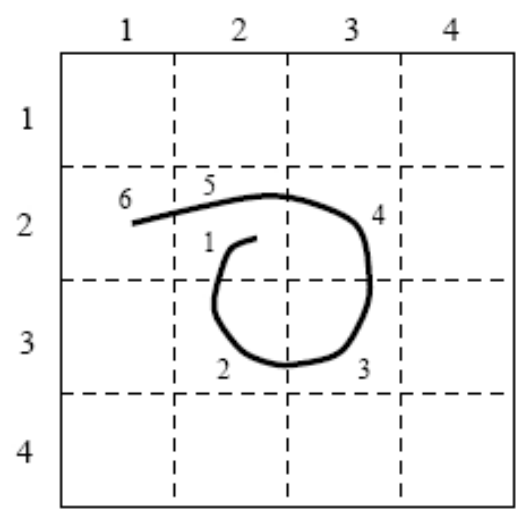

Figure 2: Draw a Secret (DAS) method on a 4*4 Grid

***Weaknesses***: Goldberg in 2002 had a survey which showed that most of the uses forgot their stroke order. On the other hand, he showed that the user can remember text password easier than DAS Password. The other weakness is that, the users tend to choose frail graphical passwords that are vulnerable to the graphical dictionary attack [11].

### 3. Grid Selection (Pure recall)

In 2004, Thorpe and van Oorschot further studied the impact of password length and stroke-count as a complexity property of the DAS scheme. Their study showed that stroke-count has the largest impact on the DAS password space -- The size of DAS password space decreases significantly with fewer strokes for a fixed password length. The length of a DAS password also has a significant impact but the impact is not as strong as the stroke-count. To improve the security, Thorpe and van Oorschot proposed a "Grid Selection" technique. The selection grid is an initially large, fine grained grid from which the user selects a drawing grid, a rectangular region to zoom in on, in which they may enter their password (figure 3). This would significantly increase the DAS password space [20].

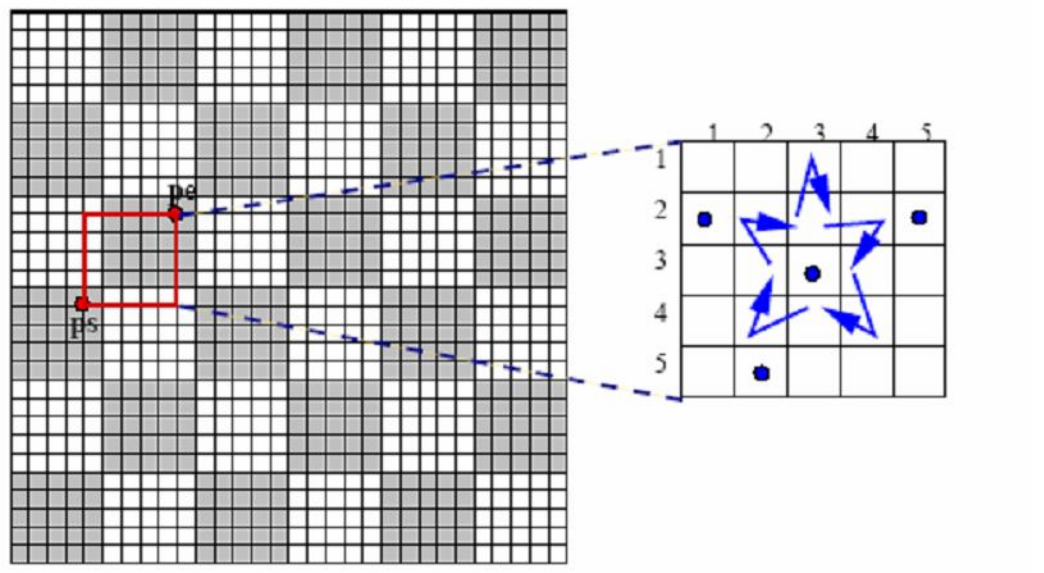

Figure 3: A sample of Grid Selection method

***Weaknesses***: This method just significantly increases the DAS password space but the lacks of DAS doesn't solve yet [20].

### 4. Qualitative DAS (QDAS) (Pure recall)

In 2007, QDAS method designed as an enhancement of DAS method created by encoding each stroke. The raw encoding consists of its starting cell and the sequence of qualitative direction change in the stroke relative to the grid. A direction change is considered when the pen cross a cell boundary in a direction different from direction the cross the previous cell boundary. The research shows that, the image which has more area of interest (Hot Spot) could be more useful as a background image [2]. Figure 4 shows a sample of QDAS password. Figure 4 shows a sample of QDAS password.

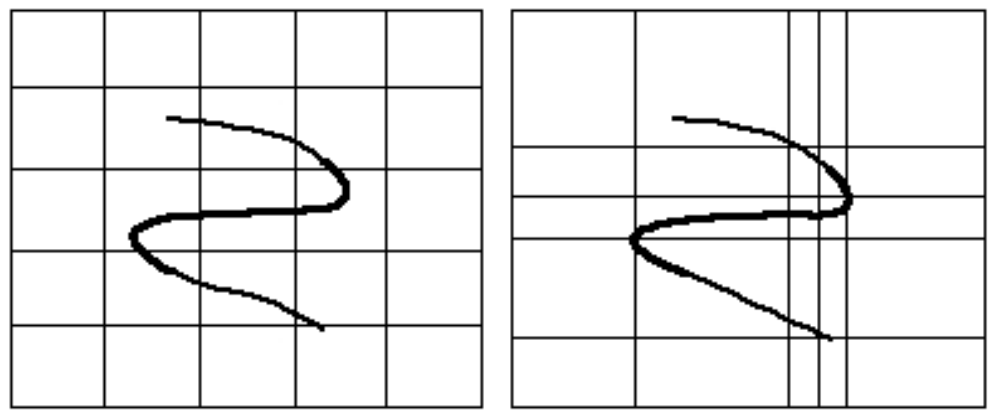

Figure 4: A sample of Qualitative DAS Algorithm

***Weaknesses***: This model uses dynamic grid transformation to hide the process of creating password so this method could be safer that original DAS to shoulder surfing attack. Although this model have more entropy than previous DAS but it has less memorable than the original one [2].

5. Syukri Algorithm (Pure recall)

Syukri algorithm proposes a system where authentication is conducted by having user drawing their signature using mouse. See Figure 5 [21]. This technique includes two stages, namely, registration and verification. During the registration stage, user will first be asked to draw their signature with mouse, and then the system will extract the signature area and either enlarges or scale-down signatures, rotates if needed, (also known as normalizing). The information will later be saved into the database. The verification stage first takes the user input, and does the normalization again, and then extracts the parameters of the signature. The system conducts verification using geometric average means and a dynamic update of database. According to the study [21], the rate of successful verification was satisfying. The biggest advantage of this approach is that there is no need to memorize one's signature and signatures are hard to fake. Figure 5 shows a sample of Syukri password.

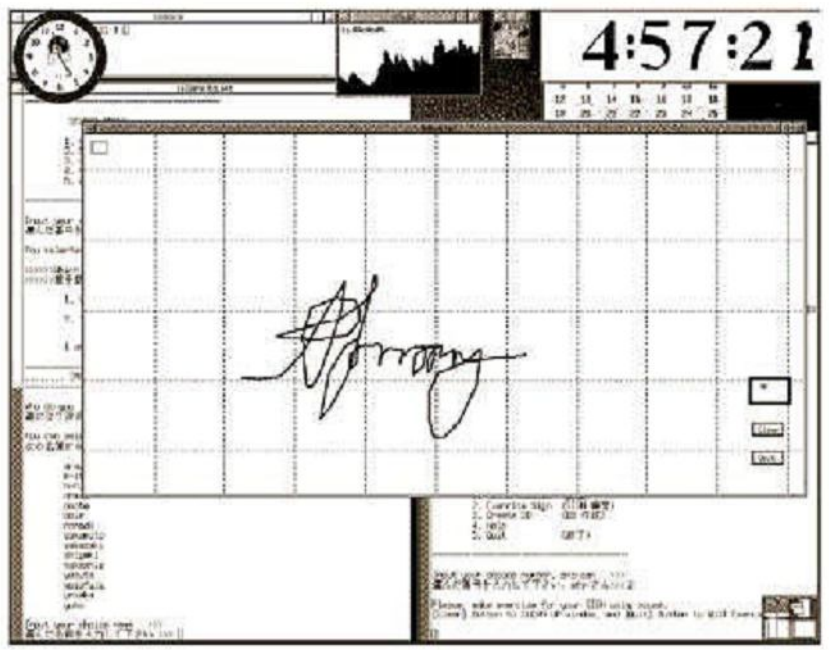

Figure 5: A sample of Syukri algorithm

***Weaknesses***: However, not everybody is familiar with using mouse as a writing device; the signature can therefore be hard to drawn. One possible solution to this problem would be to use a pen-like input device, but such devices are not widely used, and adding new hardware to the current system can be expensive. In this study, researchers believed such technique is more useful to small devices [21].

6. Blonder (cute recall)

In 1996, this method designed by Greg E. Blonder which a pre-determined image presented to the user on a visual display and user should be point to one or more predetermined positions on the image (tap regions) in a predetermined order as a way of point out his or her authorization to access the resource. Originator maintained that the method is secure according to a millions of different regions. Figure 6 shows a sample of Blonder password. 

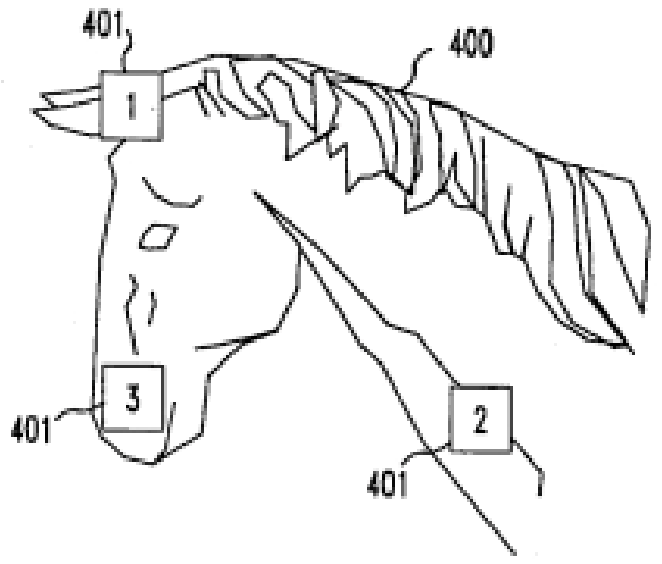

Figure 6: A sample of Blonder method

***Weaknesses***: A problem with this scheme was that the number of predefined click regions was relatively small so the password had to be quite long to be secure. Also, the use of pre-defined click objects or regions required simple, artificial images, for example cartoon-like images, instead of complex, real-world scenes [22].

7. PassPoint (cute recall)

In 2005, PassPoint created in order to cover the limitation of Blonder Algorithm which was limitation of image. The picture could be any natural picture or painting but at the same time should be rich enough in order to have many possible click points. On the other hand the image is not secret and has no role other than helping the user to remember the click point. Another source of flexibility is that there is no need for artificial predefined click regions with well-marked boundaries like blonder algorithm. The user is choosing several points on picture in a particular order [16]. Figure 7 shows a sample of PassPoint password.

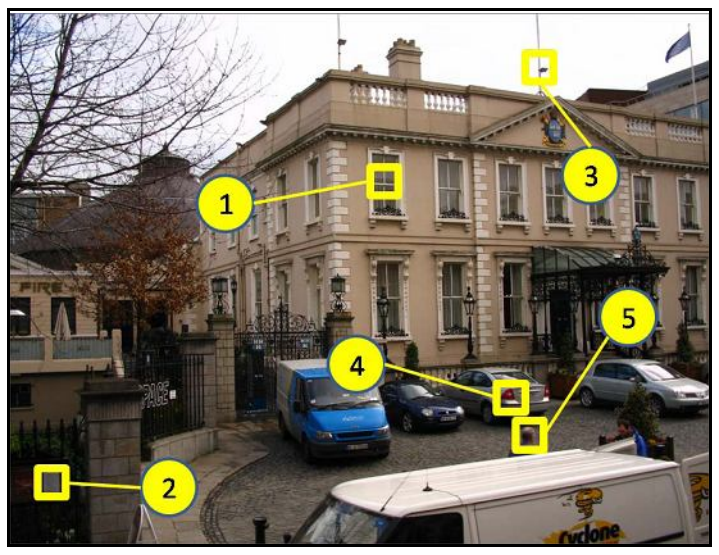

Figure 7: A sample of Passpoint method

***Weaknesses***: Users in PassPoint system were able to easily and quickly create a valid password, but they had more difficulty learning their passwords than alphanumeric users, taking more trials and more time to complete the practice, On the other hand the login time, in this method is longer than alphanumeric method [16].

8. Background DAS (BDAS) (cute recall)

In 2007, this method proposed by adding background image to the original DAS for improvement, so that both background image and the drawing grid can be used to providing cued recall [11]. The user starts by using three different ways:

i. The user have secret in mind to begin, and then draw using the point from a background image.

ii. The user's choice of secret is affected by various characteristic of the image.
iii. A mix of two above methods.

Figure 8 shows a sample of BDAS algorithm.

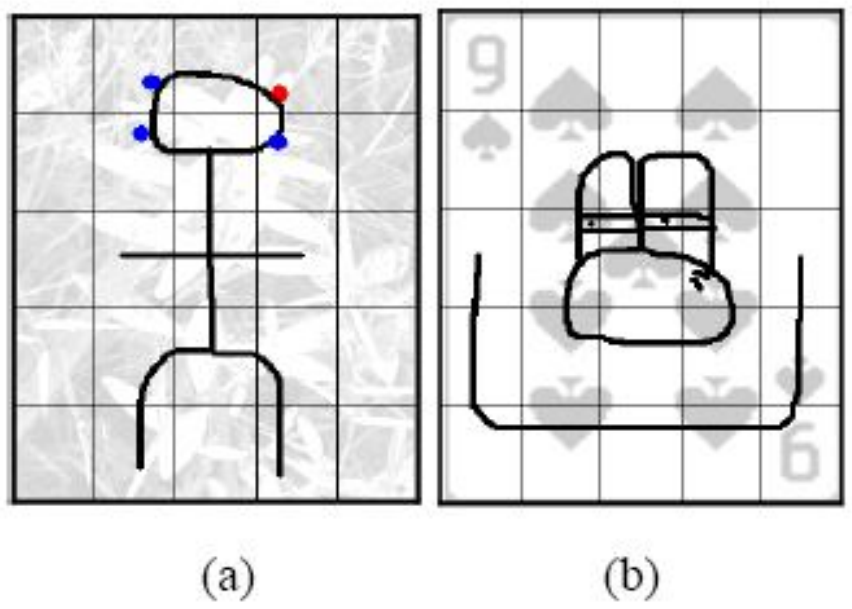

Figure 8: A sample of BDAS algorithm

***Weaknesses***: With reference to a research on BDAS, memory decaying over a week is one of the major problems in this algorithm. Users had no problem in recreating it in the five-minute test, but a week later they could not do better than producing the secret password as previous. Also shoulder-surfing and interference between multiple passwords are concerns for BDAS [11].

9. PASSMAP (cute recall)

One of the main problems with passwords is that very good passwords are hard to remember and the one which are easy to remember are too short of simple to be secured. From the studies of human memory, we know that it is relatively easy to remember landmarks on a well-known journey [19]. Figure 9 will be shows a sample of PassMap password.

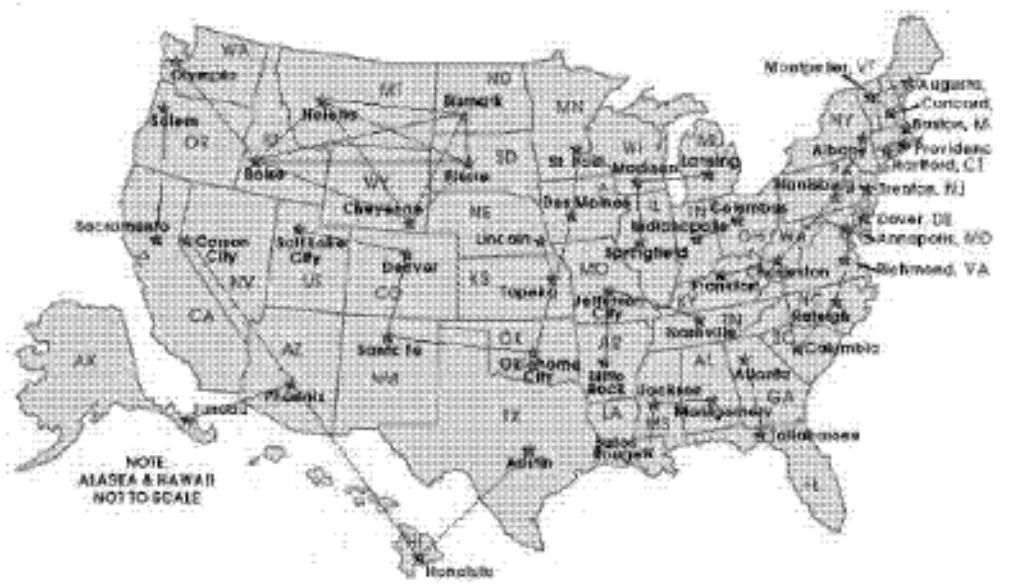

Figure 9: A sample of PASSMAP method

***Weaknesses***: Additionally the PassMap technology is not very susceptible to "shoulder surfing" as can be clearly seen from Figure 8. Noticing a single new edge in a large graph or even an absence of some edge in the map is not a trivial task, for someone just passing by. But it is respect to Brute Force attacks while at the same time considering how good those mechanisms are in terms how memorable they are [19].

10. Passlogix v-Go (cute recall)

Passlogix Inc. is a commercial security company located in New York City USA. Their scheme called Passlogix v-Go uses a technique known as "Repeating a sequence of actions" which means creating a password by a chronological situation. In this scheme, user can select their background images based on the environment, for example in the kitchen, bathroom, bedroom or others (See Figure 10). To enter a password, user can click and/or drag on a series of items within that image [20].

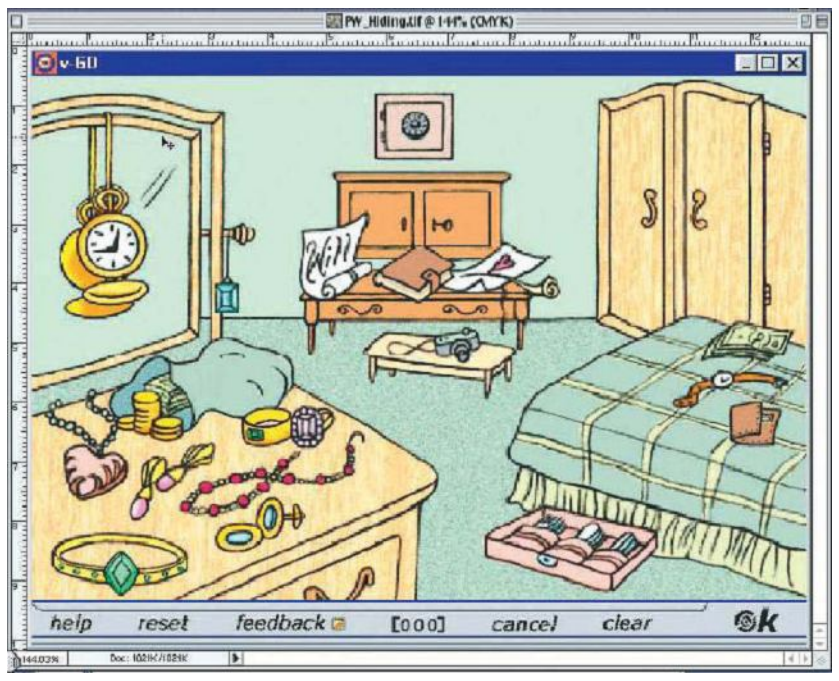

Figure 10: A sample of PASSMAP method

***Weaknesses***: There are some disadvantages such as the size of password space is small. There are limited places that one can take vegetables, fruits or food from and put into, therefore causing the passwords to be somewhat guessable or predictable [20].

11. VisKey SFR (cute recall)

VisKey is a recall-based authentication scheme that currently has been commercialized by SFR Company in Germany. This software was designed specifically for mobile devices such as PDAs. To form a password, users need to tap their spots in sequence (Figure 11) [20].

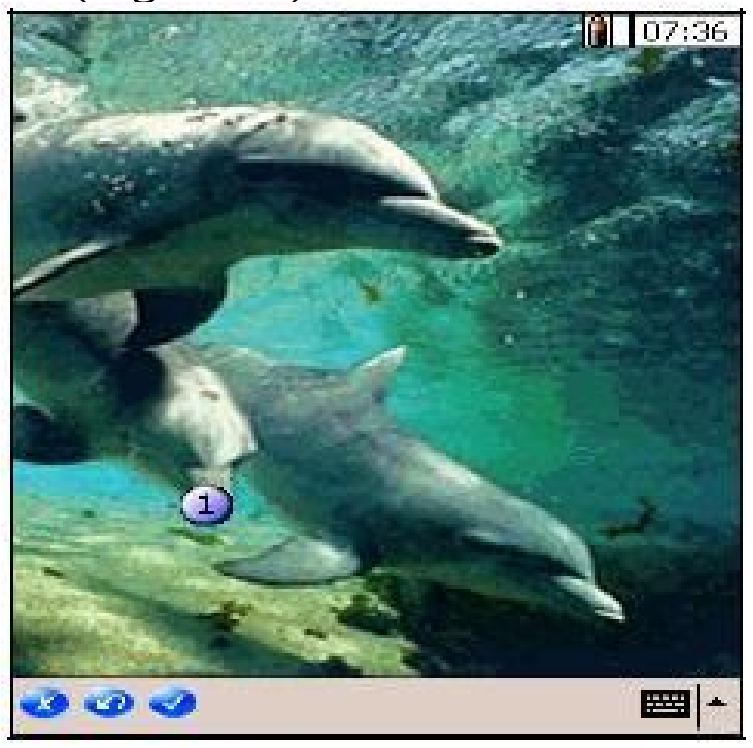

Figure 11: A sample of VisKey SFR method

***Weaknesses:*** The problem with this technique is the input tolerance. Since it is difficult to point to the exact spots on the picture, Viskey permits all input within a certain tolerance area around it. The size of this area can be pre-defined by users. Nonetheless, some precautions related to the input precision needs to be set carefully, as it will directly influence the security and the usability of the password. For a practical setting of parameters, a four spot VisKey can offer theoretically almost 1 billion possibilities to define a password. However, is not large enough to avoid the off-line attacks by a high-speed computer. At least seven defined spots are needed in order to overcome the brute force attacks [20].

12. Pass-Go Scheme

In 2006, this scheme is an improvement of DAS algorithm which kept the advantages of the DAS plus adding some extra security features to it. Pass-Go is a grid-based scheme which requires a user to select intersections, instead of cells so the new system refers to a matrix of intersections, rather than cells as in DAS. As an intersection is actually a point which doesn't have an area, it would be impossible for a user to touch it without an error tolerance mechanism. Therefore sensitive areas defined to address this problem.

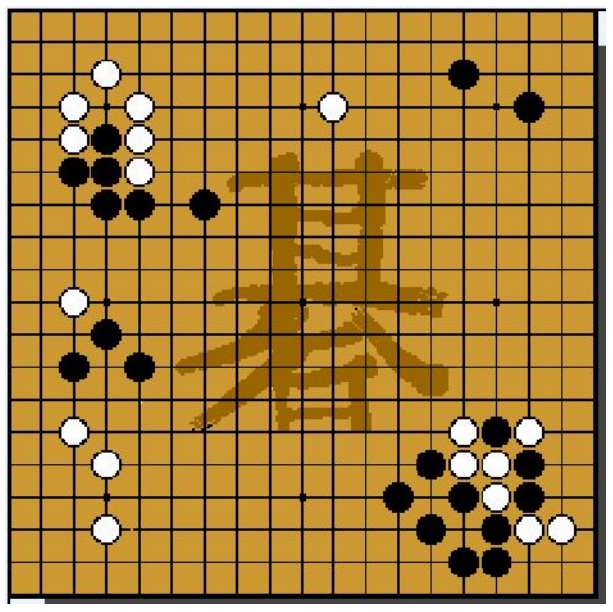

Figure 12: Pass-Go Scheme, 2006

Changing the format of typing from cell to intersection bring the user more free choices. The other difference between these two algorithms is that the size of grid in enhanced method changes to 9*9.

## III. ISO STANDARDS ON USABILITY

The International Organization for Standardization (ISO) developed a variety of models to measure usability, but none of these models cover all usability aspects. This section tries to scrutiny three models in ISO and finally makes a complete table of usability attributes base on these ISO standards.

**ISO 9241:**

ISO 9241 is a series of international standards of ergonomics requirements for office work with visual display terminals (Figure 12). The definitions of Part 11 of ISO 9241 are built from a different usability viewpoint. Its key components are: effectiveness that describes the interaction from a process point of view, efficiency that is the attention for results and resources implied and satisfaction that refers to a user point of view.

ISO 9241 provides requirements and recommendations concerning hardware, software and environment attributes that contribute to usability, and subjacent ergonomic principles. Parts 3 to 9 deal with hardware design requirements and guidelines that can have implications on software. Parts 10 to 17 deal with software attributes [25].

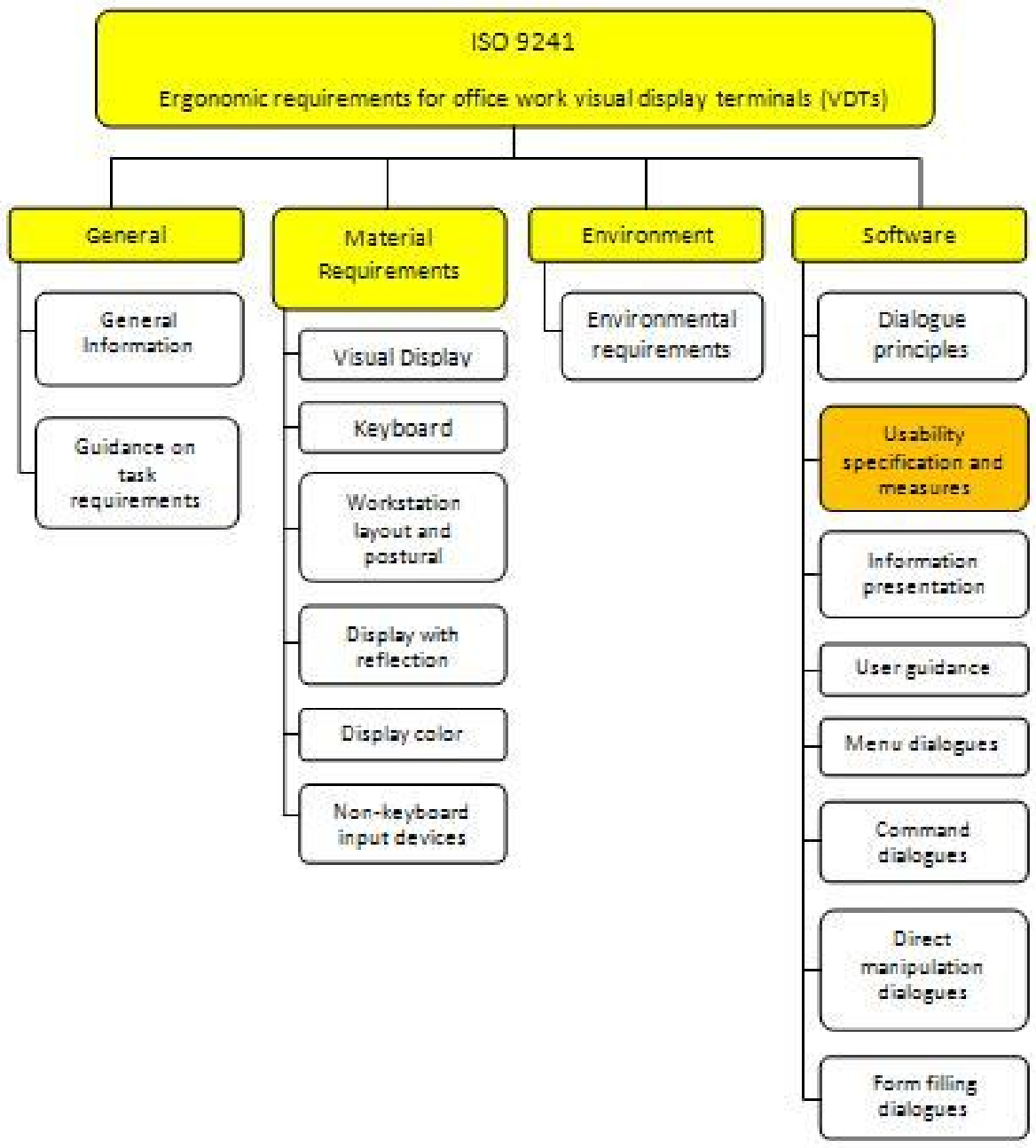

Figure 12: The 17 Parts of ISO 9241

According to this standard, the measurement of system usability consists of three usability attributes:

1. Effectiveness: How well do the users achieve the goals they set out to achieve using the system?
2. Efficiency: The resources consumed in order to achieve their goals.
3. Satisfaction: How the users feel about their use of the system?

**ISO 9126:**

ISO 9126 address software quality from the product point of view. It is probably the most extensive software quality model, even if it is not exhaustive. Initially published in 1991, the approach of its quality model is to present quality as a whole set of characteristics. It divides software quality into six general categories: functionalities, reliability, usability, effectiveness, maintainability and portability (figure 13) [25].

Part four of ISO 9126 defined the usability as "A set of attributes that bear on the effort needed for use and on the individual assessment of such use, by a stated or implied set of users". It proposed then a product oriented usability approach. Usability was seen like an independent factor of software quality. It treated software attributes, mainly its interface that makes it easy to use. As you see in figure 13, the major attributes are: Understandability, Learnability, Operability, and Attractiveness [25].

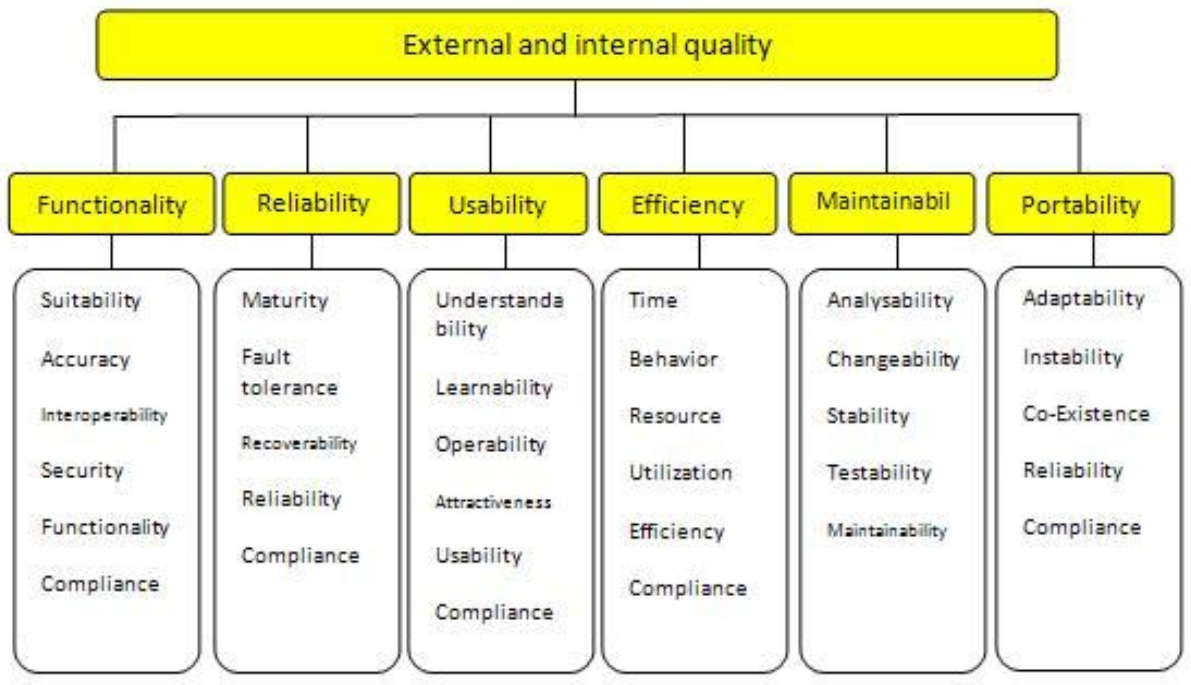

Figure 13: The 6 Parts of ISO 9126

**ISO 13407:**

This International Standard provides guidance on human-centred design activities throughout the life cycle of computer-based interactive systems. It is aimed at those managing design processes and provides guidance on sources of information and standards relevant to the human-centered approach. For the purposes of this International Standard, there are eight terms and definitions inclusive Interactive system, Prototype, Usability, Effectiveness, Efficiency, Context of use, and user that third term as usability define as below:

Extent to which a product can be used by specified users to achieve specified goals with effectiveness, efficiency and satisfaction in a specified context of use [ISO 9241-11].

Finally, The Usability Model that defined by ISO 13407 is comprised of five stages, which are implicitly joined in a loop. Figure 14 decided this model graphically [26].

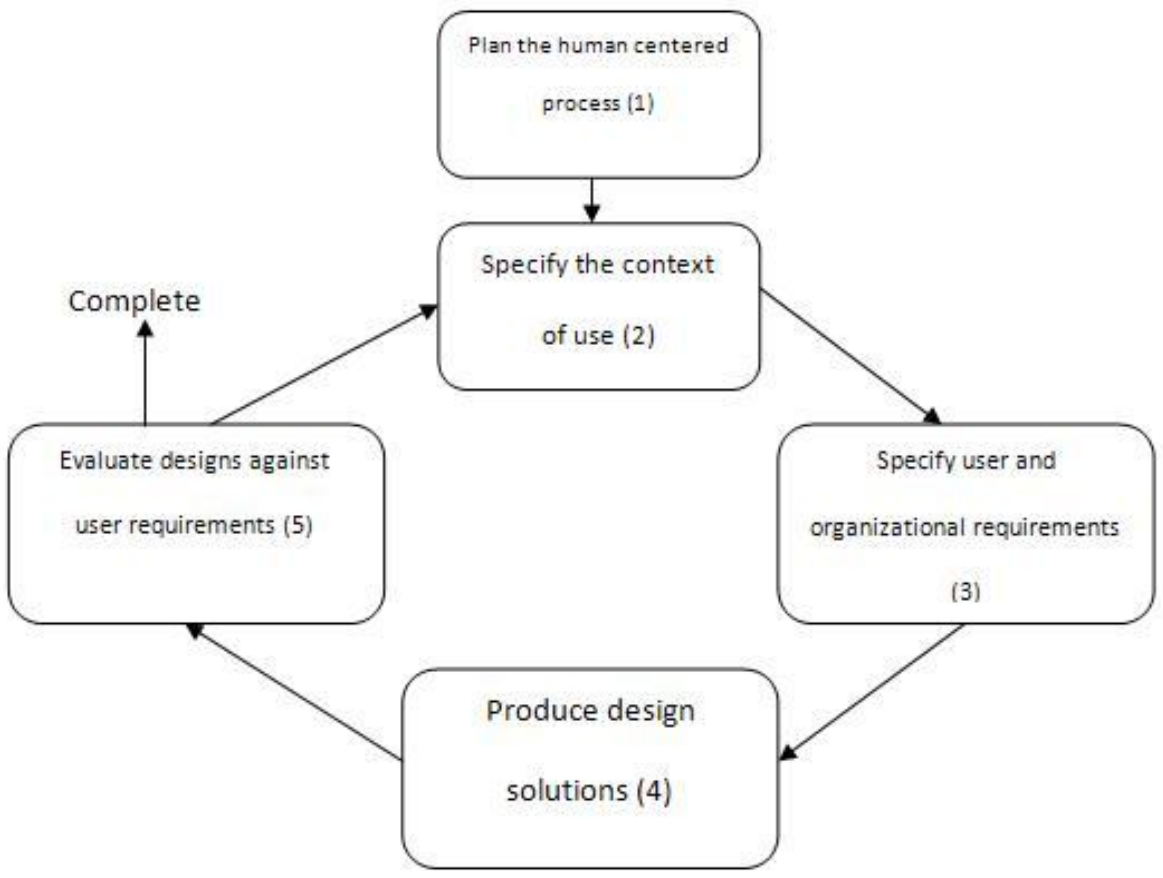

Figure 14: The 5 Parts of ISO 13407

Finally, after this survey on ISO standards (9241, 9126, 13407) in usability we find more attributes for each feature that you can see in the table below (Table 1).

Table 1: Our finalized Usability attributes from ISO Standards

| Usability features | attributes | Attributes especially for graphical user authentication | Abbreviation |
|---|---|---|---|
| Effectiveness | Reliability & Accuracy | Reliability & Accuracy | R&A |
| Efficiency | The utilization in real world | Applicable | Applicable |
| Satisfaction | Easy to use | Use the mouse easily | Mouse usage |
| | Easy to create | Select simple way to create the password | Create Simply |
| | Easy to memorize (memorability) | Meaningful | meaningful |
| | | User assign image | Assignable Image |
| | | Freedom of choice | |
| | Easy to execute | Select simple steps of registration and login | Simple Steps |
| | Good view | Select good interface | Nice interface |
| | Easy to understand | Simple training session | Training simply |
| | Pleasant | Pleasant picture | Pleasant picture |

## IV. USABILITY IN RECALL-BASED TECHNIQUES

With reference to table 1, now we can make a comparison table among all recall-based algorithms in two categories as pure and cued recall-based algorithm that you can find in tables below (Table 2, 3):

Table 2: The Usability features in Pure Recall-Based Techniques

| Row | Pure recall-based algorithm | Usability Features | | | | | | | | |
|---|---|---|---|---|---|---|---|---|---|---|
| | | Satisfaction | | | | | | | Efficiency | Effectiveness |
| | | Mouse usage | Create Simply | meaningful | Memorability | Simple Steps | Nice Interface | Training Simply | Applicable | R&A |
| 1 | Passdoodle | Y | N | Y | N | Y | NA | Y | Y | N |
| 2 | Draw A Secret(DAS) | Y | N | N | N | Y | NA | Y | Y | Y |
| 3 | Grid Selection | Y | N | N | N | Y | NA | Y | N | Y |
| 4 | Qualitative DAS | Y | N | N | N | Y | NA | Y | Y | N |
| 5 | Syukri Algorithm | Y | N | Y | Y | Y | Y | Y | Y | Y |

Table 3: The Usability features in Cued Recall-Based Techniques

| Row | Cued recall-based algorithm | Usability Features | | | | | | | | | | |
|---|---|---|---|---|---|---|---|---|---|---|---|---|
| | | Satisfaction | | | | | | | | | Efficiency | Effectiveness |
| | | Mouse usage | Create Simply | meaningful | Clickable Points | Memorability | Simple Steps | Nice Interface | Training Simply | Pleasant Picture | Applicable | R&A |
| 1 | Blonder | Y | Y | N | Y | Y | Y | N | Y | N | Y | N |
| 2 | PassPoint | Y | Y | N | Y | Y | Y | Y | Y | Y | Y | Y |
| 3 | Background DAS | Y | N | Y | N | Y | N | N | N | N | N | Y |
| 4 | PASSMAP | Y | Y | Y | Y | Y | Y | N | Y | N | Y | N |
| 5 | Passlogix v-Go | Y | N | Y | N | Y | N | Y | N | Y | Y | Y |
| 6 | VisKey SFR | Y | Y | N | Y | Y | Y | Y | Y | Y | Y | Y |
| 7 | Pass-Go | Y | Y | N | Y | Y | Y | N | Y | N | N | Y |

## V. LIST OF ATTACKS ON GUA

With reference to the Common Attack Pattern Enumeration and Classification (CAPEC) Standard Abstraction Attack Pattern List (Release 1.3) and other resources of attacks, finally we found six attacks method that is efficient in graphical user authentication (GUA) algorithms [28]. Now, this section tries to explain these attacks methods and then next section tries to make a comparison table among all recall-based algorithms that explained in section II.

### BRUTE FORCE

It is more difficult to carry out a brute force attack against graphical passwords than text-based passwords. The attack programs need to automatically generate accurate mouse motion to imitate human input, which is particularly difficult for recall based graphical passwords. Overall, we believe a graphical password is less vulnerable to brute force attacks than a text-based password. The main defense against brute force search is to have a sufficiently large password space. The speed which an attacker discovers a secret is directly related to the resources that the attacker has. This attack method is resource expensive as the attackers' chance for finding user's password is high only if the resources be as complete as possible.

### DICTIONARY ATTACKS

Since recognition based graphical passwords involve mouse input instead of keyboard input, it will be impractical to carry out dictionary attacks against this type of graphical passwords. For some recall based graphical passwords, it is possible to use a dictionary attack but an automated dictionary attack will be much more complex than a text based dictionary attack. More research is needed in this area. Overall, we believe graphical passwords are less vulnerable to dictionary attacks than text-based passwords.

### GUESSING

Password guessing attacks can be broadly categorised into online password guessing attacks and offline dictionary attacks. In an online password guessing attack, an attacker tries a guessed password by manipulating the inputs of one or more oracles. In an offline dictionary attack, an attacker exhaustively searches for the password by manipulating the inputs of one or more oracles. As many users try to select their password based on their personal information like the name of their pets, passport number, family name and so on, the attacker try to guess.

### SPYWARE

Spyware is a type of malware which installed on computers with the aim of collecting sensitive information of users, using a key logger or key listener. This information gathered without user's knowledge and report back to an outside source. During graphical password authentication the attacker attempt to gain sensitive information like user names or selected passwords images by intercepting information exchanged.

### SHOULDER SURFING

Shoulder surfing refers to using direct observation techniques, such as looking over someone's shoulder, to get information. Shoulder surfing is effective in crowded places because it's really easy to stand near someone and watch them entering a PIN number at an ATM machine. This attack is also possible at a distance using vision-enhancing devices like miniature closed-circuit television cameras can be concealed in ceilings, walls or fixtures to observe data entry. To prevent shoulder surfing, it is advised to shield paperwork or the keypad from view by using one's body or cupping one's hand.

### SOCIAL ENGINEERING (DESCRIPTION)

In this kind of attack an attacker uses human interaction to obtain or compromise information about an organization or computer systems. An attacker possibly claiming to be a new employee, or researcher and even offering credentials to support that identity. However, by asking questions, he or she may be able to piece together enough information to infiltrate an organization's network. If an attacker is not able to gather enough information from one source, he or she may contact another source within the same organization and rely on the information from the first source to add to his or her credibility.

## VI. IMPRESSIBILITY ON ATTACKS IN RECALL-BASED ALGORITHMS

Previous part tries to explain and define six major attack patterns in graphical user authentication technology. Now, this part makes a comparison table among all recall-based algorithms based on attack patterns and impressibility of algorithms on attack patterns.

Table 4: The attacks peruse in Recall-Based algorithms

| Row | Algorithm | Cued Recall-Based | Pure Recall-Based | Attacks | | | | | |
|---|---|---|---|---|---|---|---|---|---|
| | | | | Brute Force | Dictionary | Guessing | Spyware | Shoulder Surfing | Social Engineering |
| 1 | Passdoodle | • | | N | | | | | |
| 2 | Draw A Secret(DAS) | • | | N | Y | Y | N | Y | N |
| 3 | Grid Selection | • | | N | | | | | |
| 4 | Qualitative DAS | • | | N | | | | | |
| 5 | Syukri Algorithm | • | | N | Y | Y | N | Y | N |
| 6 | Blonder | | • | Y | N | Y | N | Y | N |
| 7 | PassPoint | | • | Y | N | Y | N | Y | N |
| 8 | Background DAS | | • | N | | | | | |
| 9 | PASSMAP | | • | Y | N | | N | Y | N |
| 10 | Passlogix v-Go | | • | Y | N | Y | N | Y | N |
| 11 | VisKey SFR | | • | Y | N | Y | N | Y | N |
| 12 | Pass-Go | | • | Y | | | | | |

## VII. CONCLUSIONS

In this study, twelve algorithms from Recall-Based explained in two Pure and Cued categories and peruse most of their weaknesses and vulnerabilities. In the first part of research, we found three ISO standard included ISO 9241, ISO 9126 and ISO 13407 and find three main usability categories and then for each category define some sub attributes that you can find in table1. Then, tables 2 and table 3 showed the comparison among all pure and cued recall-based algorithms on our useability founded attributes and sub attributes. In the second part, we found the effective attack patterns on graphical user authentication (GUA) and explained them. Finally, in the last part we made a comparison table among impressibility of all recall-based algorithms based on standard attack patterns.

## ACKNOWLEDGMENT

We would like to express our appreciation to our parents and all the teachers and lecturers who help us to understand the importance of knowledge and show us the best way to gain it.

## REFERENCES

[1] Ahmet Emir Dirik, Nasir Memon and Jean-Camille Birget, "Modeling user choice in the PassPoints graphical password scheme", Symposium on Usable Privacy and Security 2007. Pittsburgh, Pennsylvania, USA. ACM. 20-28; July 2007.

[2] Di Lin, Paul Dunphy, Patrick Olivier and Jeff Yan, "Graphical Passwords & Qualitative Spatial Relations", Proceedings of the 3rd symposium on Usable privacy and security. Pittsburgh, Pennsylvania. ACM. 161-162 ; July 2007.

[3] Eiji Hayashi , Nicolas Christin, "Use Your Illusion: Secure Authentication Usable Anywhere", Proceedings of the 4th symposium on Usable privacy and security (SOUPS). Pittsburgh, PA USA, ACM. 35-45, July 2008.

[4] Furkan T., A. Ant Ozok, and Stephen H. Holden, "A Comparison of Perceived and Real Shoulder-surfing Risks between Alphanumeric and Graphical Passwords", Symposium on Usable Privacy and Security (SOUPS). Pittsburgh, Pennsylvania, USA. ACM. 56-66; July 2006.

[5] Greg E. Blonder , Graphical Password U.S. Patent No. 5559961, 1996.

[6] Haichang Gao, Xuewu Guo, Xiaoping Chen, Liming Wang, and Xiyang Liu, "YAGP: Yet another Graphical Password Strategy", 2008 Annual Computer Security Applications Conference. IEEE. 121-129; 2008.

[7] Jermyn Ian, A. Mayer, F. Monrose, M. K. Reiter and A. D. Rubin,"The design and analysis of graphical passwords", Proceedings of the Eighth USENIX Security Symposium. August 23-26 1999. USENIX Association 1–14, 1999.

[8] Julie Thorpe, P.C. van Oorschot,"Towards Secure Design Choices for Implementing Graphical Passwords", Proceedings of the 20th Annual Computer Security Applications Conference. Ottawa, Ont., Canada, IEEE. 50 – 60; Dec 2004.

[9] L. Y. POR, X. T. LIM; "Multi-Grid Background Pass-Go", WSEAS TRANSACTIONS on INFORMATION SCIENCE & APPLICATIONS, ISSN: 1790-0832, Issue 7, Volume 5, July 2008.

[10] Paul Dunphy , James Nicholson , Patrick Olivier, "Securing Passfaces for Description", Symposium on Usable Privacy and Security (SOUPS), Pittsburgh, PA USA; 2008.

[11] Paul Dunphy, Jeff Yan, "Do Background Images Improve "Draw a Secret" Graphical Passwords?", Proceedings of the 14th ACM conference on Computer and communications security. Alexandria, Virginia, USA. ACM. 36-47; 2007.

[12] Rachna Dhamija, "Hash visualization in user authentication", Proceedings of CHI 2000. The Hague, the Netherlands. ACM 279–280; 2000.

[13] Roman Weiss, Alexander De Luca, "PassShapes – Utilizing Stroke Based Authentication to Increase Password Memorability", Proceedings of the 5th Nordic conference on Human-computer interaction: building bridges. Lund, Sweden. ACM. 2008. 383-392; October 2008.

[14] Sacha Brostoff & M. Angela Sasse, "Are Passfaces1 More Usable Than Passwords? (A Field Trial Investigation)", Department of Computer Science, University College London,London, WC1E 6BT; 2008.

[15] Saranga Komanduri, Dugald R. Hutchings, "Order and Entropy in Picture Passwords", Proceedings of graphics interface 2008. Windsor, Ontario, Canada. Canadian Information Processing Society. 115-122; May 2008.

[16] Susan Wiedenbecka, Jim Watersa, Jean-Camille Birgetb and Alex Brodskiyc, Nasir Memon. PassPoints, "Design and longitudinal evaluation of a graphical password system", Academic Press, Inc. 102-127, July 2005

[17] Xiaoyuan Suo, Ying Zhu and G. Scott. Owen, "Graphical Passwords: A Survey", Proceedings of the 21st Annual Computer Security Applications. IEEE. 463-472; 2005.

[18] Christopher Varenhorst," Passdoodles; a Lightweight Authentication Method " , Massachusetts Institute of Technology, Research Science Institute, July 27,2004.

[19] Roman V. Yampolskiy, "User Authentication via Behavior Based Passwords"; IEEE Explore, 2007.

[20] Muhammad Daniel Hafiz, Abdul Hanan Abdullah, Norafida Ithnin, Hazinah K. Mammi; "Towards Identifying Usability and Security Features of Graphical Password in Knowledge Based Authentication Technique"; IEEE Explore, 2008.

[21] Ali Mohamed Eljetlawi, "Study and Develop a New Graphical Password System", University Technology Malaysia, Master Dissertation, 2008.

[22] Susan Wiedenbeck, Jean-Camille Birget, Alex Brodskiy;" Authentication Using Graphical Passwords:Effects of Tolerance and Image Choice", Symposium On Usable Privacy and Security (SOUPS), Pittsburgh, PA, USA, 2005.

[23] Paul Dunphy, Jeff Yan; "Do Background Images Improve "Draw a Secret" Graphical Passwords?"; CCS'07, Alexandria, Virginia, USA, 2007.

[24] Karen Renaud, "On user involvement in production of images used in visual authentication" ; Elsevier, Journal of Visual Languages and Computing,2008.

[25] Alain Abran, Witold Suryn, Adel Khelifi, Juergen Rilling, Ahmed Seffah; “Consolidating the ISO Usability Models”, Concordia University, Montreal, Canada

[26] International Standard ISO 13407, “Human-centred design processes forinteractive systems”, First edition, 1999-06-01

[27] Common Attack Pattern Enumeration and Classification (CAPEC) Standard Abstraction Attack Pattern List (Release 1.3); http://capec.mitre.org/data/lists/patabs_standard.html, Access on October 2009.